\documentclass[twocolumn]{aastex631}

\begin{document}

\title{uGMRT and MeerKAT observation of RXCJ0232-4420: a quiet cluster with a giant radio halo}

\author[0009-0002-2173-0953]{Pralay Biswas}
\affiliation{National Centre for Radio Astrophysics, Tata Institute of Fundamental Research, Post Bag 3, Ganeshkhind, Pune 411007, Maharashtra, India}

\author[0009-0002-0373-570X]{Ramananda Santra}
\affiliation{National Centre for Radio Astrophysics, Tata Institute of Fundamental Research, Post Bag 3, Ganeshkhind, Pune 411007, Maharashtra, India}
\affiliation{International Centre for Theoretical Sciences, Tata Institute of Fundamental Research, Bangalore 560089, India}

\author[0000-0003-1449-3718]{Ruta Kale}
\affiliation{National Centre for Radio Astrophysics, Tata Institute of Fundamental Research, Post Bag 3, Ganeshkhind, Pune 411007, Maharashtra, India}

\author[0000-0001-6282-6025]{Viral Parekh}
\affiliation{National Radio Astronomy Observatory (NRAO), 1003 Lopezville Rd, Socorro, NM 87801, USA}



\begin{abstract}
Giant radio halos are Mpc-scale extended sources associated with merging clusters, while minihalos are preferentially associated with cool-core clusters. Both trace the intracluster medium (ICM) plasma physical process, and recent low-frequency observations increasingly blur the distinction between the two classes. We present the first multifrequency spectral analysis of the galaxy cluster RXCJ0232--4420, which hosts a cool core, using uGMRT (400 and 650 MHz) and MeerKAT (1283 MHz) observations. The central radio emission extends beyond $\sim 1$ Mpc at all frequencies, confirming it as a giant radio halo. One candidate relic (in the east) has also been detected, with an extent of $\sim 300$ kpc. The integrated spectral indices of halo and candidate east relic are $\alpha = -1.17 \pm 0.17$, and $\alpha = -0.85 \pm 0.17$, respectively. The resolved spectral map of the halo is mostly uniform ($-1.0$ to $-1.3$) and does not show any radial steepening. The radio surface brightness profile is well modelled by a single exponential law, with the e-folding radius constant across frequencies. The radio halo emission is morphologically well correlated with the thermal emission. Point-to-point radio-X-ray correlation analysis gives a sublinear relationship (slope $\sim 0.80$), with no frequency evolution. The presence of Mpc-scale emission in the cool-core cluster shows that such emission can arise in dynamically intermediate systems. Our results demonstrate that merger-driven turbulence, even from minor disturbances, can sustain cluster-wide particle reacceleration without destroying the cool core. 
\end{abstract}




\section{Introduction} \label{sec:intro}

Mergers of galaxy clusters release enormous amounts of energy ($\sim 10^{64}$ erg), driving shocks and turbulence that heat and disturb the intracluster medium (ICM) \citep[e.g.,][]{forman1982, Sarazin2002, Markevitch2007}. Radio observations have revealed an increasing number of clusters hosting extended, diffuse synchrotron emission on Mpc scales \citep[see][for reviews]{Feretti2012, Weeren2019, paul2023}. This emission provides strong evidence for magnetic fields ($\mu$G) and ultrarelativistic cosmic-ray electrons ($\gamma \sim 10^3$) permeating the cluster volume \citep[e.g.,][]{Cuciti2022, Salunkhe2025, rajpurohit25}, in some cases out to the virial radius \citep[e.g.,][]{Botteon2022b}. Due to their short radiative lifetimes, dominated by synchrotron and inverse Compton (IC) losses, these electrons cannot propagate over cluster-wide scales and must therefore be continuously generated or re-accelerated in situ within the ICM \citep[e.g.,][]{jaffe77, Brunetti2014}. Diffuse radio emission on scales larger than 100 kpc is commonly classified into three categories: giant radio halos (GRHs), minihalos (MHs), and radio relics \citep[e.g.,][]{Giovannini1999, Feretti2012}.

Radio relics typically reside on the outskirts of merging clusters and differ from radio halos in their elongated shape. These relics range in size from $\sim100$ kpc to $\sim2$ Mpc and are characterised by high degrees of linear polarisation, reaching up to 30\% \citep{Giovannini2009, Weeren2010, Kierdorf2017, Gennaro2018, Gasperin2022, Pal2025b}. In some systems, relics appear as symmetric pairs on opposite sides of the cluster, roughly perpendicular to the merger axis \citep{Bagchi2011, Weeren2011a, Koribalski2024, rajpurohit25}. Their peripheral location, elongated morphology, and strong polarisation collectively support a scenario in which radio relics trace shock fronts generated during cluster mergers \citep{Ensslin1998, Hoeft2007, Weeren2010, Rajpurohit2018, Pal2025a}. This shock-acceleration scenario is further strengthened by the spatial coincidence of such relics with discontinuities in X-ray surface brightness and temperature \citep[e.g.,][]{Finoguenov2010, Akamatsu2013}.

Conventionally, GRHs have been characterised by a smooth and regular morphology cospatial with the ICM thermal emission \citep[see][for reviews]{Weeren2019}. However, high-sensitivity observations with current instruments are increasingly revealing complex internal substructures, such as filaments and sharp edges, within these sources \citep[e.g.,][]{Botteon2022a, Rajpurohit2023, Balboni2026}. GRHs, with an extent of $1-2$ Mpc, are typically observed in massive, dynamically disturbed, merging clusters \citep{Giovannini1999, Buote2001, Cassano2010, kale15, Cuciti2021}. The largest sample of radio halos to date, compiled from the LOFAR Two-metre Sky Survey (LoTSS; \citealt{shimwell22}) at 144 MHz, reveals that giant radio halos are present in about $\sim 30 \%$ of galaxy clusters \citep{Botteon2022a}. The formation of GRHs can be explained by the turbulent reacceleration model, which results from reacceleration of the seed electron population \citep{Brunetti2001, vazza24} due to merger-driven turbulence \citep{Brunetti2001, Brunetti2007, Donnert2013, Brunetti2016}. An alternative hadronic (secondary) origin for radio halos has been proposed, in which cosmic-ray electrons are produced via proton–proton collisions in the ICM \citep[e.g.,][]{Dennison1980, Blasi1999, Dolag2000, Miniati2001, Keshet2010}. However, the observed halo properties and the lack of associated $\gamma$-ray emission in Fermi-LAT observations disfavour a purely hadronic scenario \citep[e.g.,][]{ackermann2010, ackermann2014, Brunetti2017, manna2024}.

Radio minihalos are diffuse synchrotron sources located in the cores of relaxed, cool-core galaxy clusters, with typical extents of $\sim 200–400$ kpc \citep[e.g.,][]{giacintucci14, Giacintucci2017}. They are commonly interpreted as arising from the turbulent reacceleration of cosmic-ray electrons driven by sloshing motions in the cluster core \citep[e.g.,][]{Gitti2002, zuhone11, zuhone13, machado15}, although a contribution from secondary electrons produced in hadronic collisions has also been proposed \citep[e.g.,][]{Pfrommer2004, Keshet2010, Ignesti2020}. However, recent low-frequency studies have revealed that minihalos can coexist with giant radio halos within the same system, likely tracing mergers that preserve the core while injecting turbulence at larger radii \citep[e.g.,][]{venturi17, Savini2018, Bruno2023, Weeren2024, riseley24}. Systematic LOFAR observations further show that diffuse cluster-scale emission is strongly associated with the presence of cold fronts, often exhibiting asymmetric morphologies that follow the X-ray surface brightness distribution \citep[e.g.,][]{mazzotta08, Riseley2022a, Biava2024}. In several cases, emission beyond the core appears spatially confined by outer cold fronts, suggesting that large-scale sloshing may regulate both the extent and detectability of diffuse radio emission \citep[e.g.,][]{giacintucci20, botteon25}.

These sensitive observations raise the question of whether our understanding of diffuse radio emission in cool-core clusters is complete \citep[e.g.,][]{vanweeren2026}. However, multiple challenges limit the detectability of extended emission in systems with strong cool cores. In relaxed clusters, the sharply peaked gas density profiles imply that radio surface brightness should decline rapidly with radius, following the well-established correlation between X-ray and radio emission \citep[e.g.,][]{Govoni2001, Ignesti2020, Lusetti2024, Balboni2026}. As a consequence, faint large-scale emission may fall below current sensitivity limits. In addition, if extended halo components in cool-core clusters possess very steep radio spectra, they would be preferentially detectable only at low frequencies, further contributing to their apparent scarcity in existing observations \citep[e.g.,][]{Biava2021, Biava2024}.

In this work, we present multifrequency radio observations of the galaxy cluster RXCJ0232--4420 (hereafter RXCJ0232). The cluster was discovered in the ROSAT all-sky survey \citep{Cruddace2002} and hosts two brightest cluster galaxies (BCG-A and BCG-B) separated by $\sim 100$ kpc \citep{Pierini2008, Kale2019}. A detailed X-ray analysis by \citet{Parekh2021} classified RXCJ0232 as a relaxed cool-core cluster. Previous radio studies revealed diffuse emission centred on BCG-A, where the Legacy Giant Metrewave Radio Telescope (LGMRT) 606 MHz image ($\sigma_{\rm rms}=0.1$ mJy beam$^{-1}$) shows radio emission coincident with the X-ray peak and resembling a typical minihalo \citep{Kale2019}. Deeper observations from the MeerKAT Galaxy Cluster Legacy Survey \citep[MGCLS\footnote{\url{https://mgcls.sarao.ac.za/}};][]{Knowles2022} further reveal cluster-scale diffuse emission extending up to $ \sim 1.093$ Mpc, consistent with the size of a giant radio halo \citep{Kale2022}. In addition, the MeerKAT 1283 MHz image shows two candidate relics located east and south of the diffuse emission. Together, these properties identify RXCJ0232 as a rare system between minihalos and giant radio halos, providing a unique opportunity to probe their physical connection. 

The paper is organised as follows. We present the radio observations and data analysis in section \ref{sec: data}. Section \ref{sec: result} describes our results regarding the continuum images, spectral analysis, and the X-ray radio correlation study. The main results are discussed in the section \ref{sec: discussion}. Finally, we summarise the whole work and list the conclusions in section \ref{sec: summary}. Throughout, we have considered a $\Lambda$CDM cosmology with H$_0 = 71$ Km s$^{-1}$ Mpc$^{-1}$, $\Omega_\mathrm{M} = 0.27$, and $\Omega_{\Lambda} = 0.73$. The redshift of cluster RXCJ0232 is $0.2836$. At this redshift, $1^{\prime\prime}$ corresponds to 4.24 kpc, and the luminosity distance (D$_\mathrm{L}$) is 1444 Mpc.

\begin{figure}
    \centering
    \includegraphics[width = 0.49\textwidth]{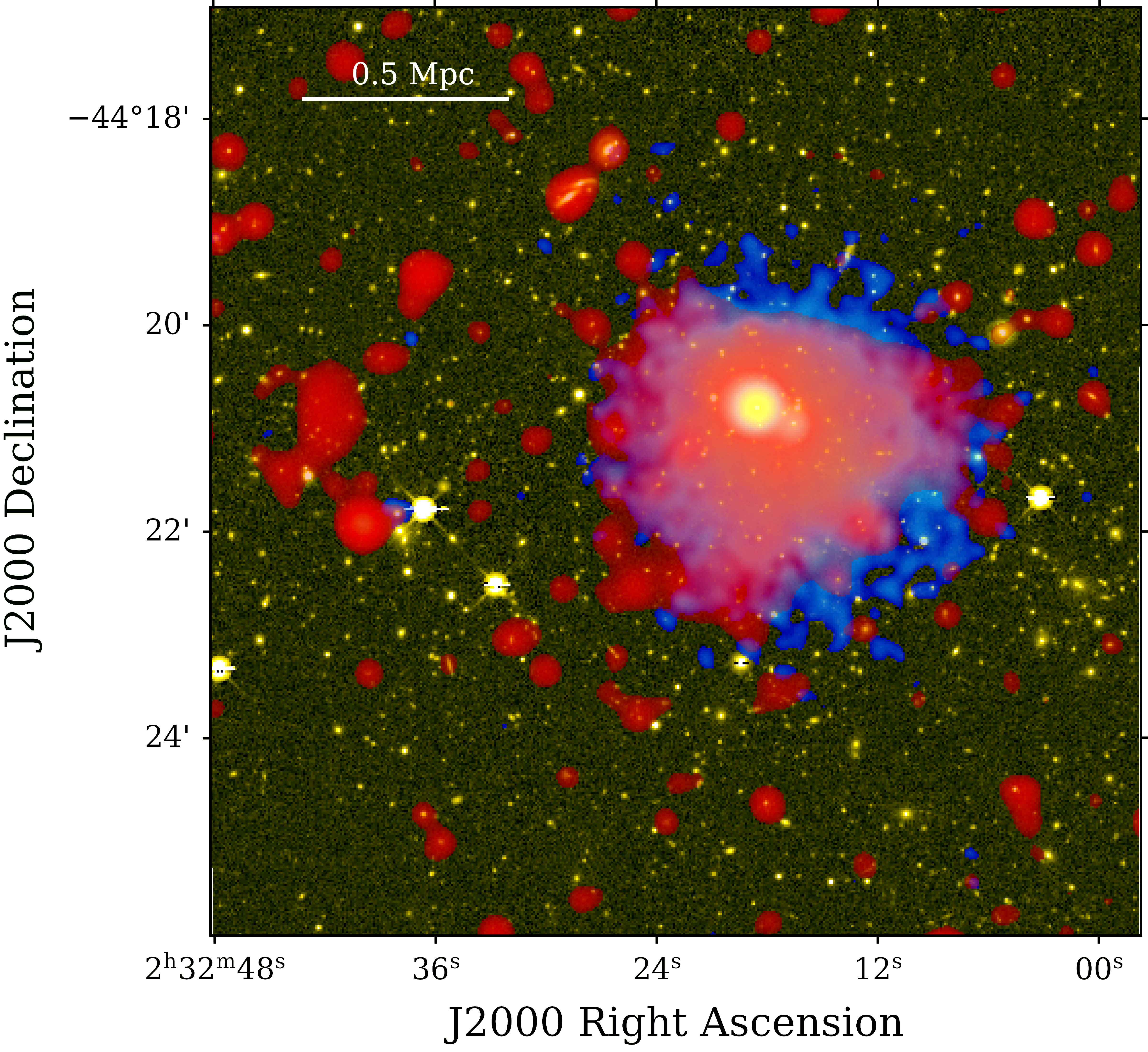}
    \caption{Multiwavelength composite image of the cluster RXCJ0232. Optical emission (yellow) shows the r-band data from the DESI Legacy Imaging Survey \citep{Dey2019}. X-ray emission (blue) displays the background-subtracted and exposure-corrected Chandra observation, smoothed with a 3-pixel Gaussian kernel, from \citet{Parekh2021}. The MeerKAT 1283 MHz radio continuum emission \citep{Knowles2022, Kale2022} is shown in red.}
    \label{fig: rgb}
\end{figure}

\section{Observation and Data Reduction} \label{sec: data}

\subsection{uGMRT}

The observations of the cluster RXCJ0232  were done using upgraded GMRT (uGMRT) Wideband Backend \citep[GWB;][]{Reddy2017} in bands 3 and 4 (central frequencies 400 and 650 MHz). The bandwidth of the observations in bands 3 and 4 is 200 MHz, spreading over 2048 channels ($36\_013$ PI: R. Kale). A summary of the observations is presented in Table \ref{table1}. At the time of observations, we used the real-time Radio Frequency Interference (RFI) filtering \citep{Buch2019, Buch2022, Buch2023} for mitigating broadband RFI.

We used \texttt{CASA} Pipeline-cum-Toolkit for Upgraded GMRT data REduction \citep[\texttt{CAPTURE}\footnote{\url{https://github.com/ruta-k/CAPTURE-CASA6}};][]{Kale2021} based on Common Astronomy Software Applications (\texttt{CASA}). First, we performed the standard flagging and calibration (complex gain and bandpass). For the absolute flux density calibration, we used the Perley-Butler 2017 \citep{Perley2017} flux density scale for the primary calibrator. After that, the target-source-calibrated data were split and further flagged using the automated flagging tasks in \texttt{CASA}. The data were averaged over 20 frequency channels ($\sim 976 \rm \ kHz$) to reduce the volume for imaging. The target measurement set was subdivided into eight subbands, each having 20 frequency channels. We use the \texttt{tclean} task in \texttt{CASA} to image the target visibilities. Several rounds of phase-only and phase \& amplitude self-calibration were done to improve the dynamic range of the images. An identical procedure is followed for both bands. We used the `Briggs' \citep{Briggs1995} weighting scheme with \texttt{robust} parameter = 0 and multiterm multifrequency synthesis (mtmfs), with \texttt{nterms=2}.


To properly detect diffuse radio emission, we subtract the discrete point sources from the full-resolution image. This process enables accurate measurement of the flux density for the diffuse emission. First, we create an image of the point sources using uv baselines greater than 5k$\lambda$ ($\sim 40^{\prime\prime}$) using \texttt{tclean}, and subtract the corresponding model visibility from the original visibility using \texttt{uvsub}. We then image the residual visibility by running \texttt{tclean} with uv-range 0.2$-$10k$\lambda$. For this, we used the `mtmfs' \texttt{deconvolver} with \texttt{robust} parameter 0. The same procedure is used for both bands to match the spatial scales of the images, which is essential for analysing the resolved spectral index map. All the final images are corrected for the primary beam response using the task \texttt{ugmrtpb}\footnote{\url{https://github.com/ruta-k/uGMRTprimarybeam-CASA6}}.

During the processing of the band 4 observation, we identified flux density offsets between the shorter and longer baselines of the uGMRT array. This is a known technical issue with the GWB backend during the 2019 observation period (see Appendix \ref{issue}). Consequently, most of the data from the central square had to be flagged, which resulted in a reduced recovery of the total flux density and extent of the diffuse emission at 650 MHz.

\begin{table}
  \centering
  \caption{Summary of uGMRT observations.} 
  \begin{tabular}{@{}lcccc@{}}
    \hline \hline
         & Date        & Frequency & Central   & On source \\
    Band & of          & range     & frequency & time      \\
         & observation & (MHz)     & (MHz)     & (hour)    \\
    \hline \hline
    band--3      & 10th Aug 2019   & $300-500$ & 400       & 2.0  \\
    band--4      & 9th Aug 2019    & $550-750$ & 650       & 1.5   \\
   \hline
  \end{tabular}
  \label{table1}
\end{table}

\subsection{MeerKAT}
MeerKAT, located in the Karoo desert of South Africa, is one of the precursors of the Square Kilometre Array. It is a next-generation radio interferometer with 64 dishes, each having a 13.5 m diameter and a largest baseline of 8 km \citep{Jonas2009}. Here, we use data from the MGCLS \citep{Knowles2022}. It is a large radio survey of 115 galaxy clusters using the MeerKAT L-band ($900-1670$ MHz) receiver. The observations have full polarisation information with 4k frequency channels across $856-1712$ MHz and 8 s of integration time. Our galaxy cluster RXCJ0232 is a part of the MGCLS.

We used the fully automated MeerKAT pipeline, CARACal, for data reduction \citep{Jozsa2020}. CARACal integrates publicly available radio data reduction tools along with custom-made software fully controlled via a YAML parameter file. The default pipeline configuration includes flagging, cross-, and self-calibration processes and delivers calibrated visibilities, continuum images (for spectroscopic observations, it also provides cubes and moment maps), and diagnostics plots. For flagging, we first applied static masks to RFI bands ($856-880$ MHz, $1658-1800$ MHz, $1419.8-1421.3$ MHz), then flagged other possible RFIs with the autoflagger tricolour for the calibrators and with AOFlagger \citep{Offringa2012} for the target, using MeerKAT-optimised strategies. Cross-calibration was done by deriving delay, bandpass, and gain solutions with the usual CASA tasks. After this, we averaged the target field data using 10 frequency channels. Imaging was performed using WSCLEAN \citep{Offringa2014} with multiscale and wideband deconvolution, generating multifrequency synthesis (MFS) maps centred at 1283 MHz. Automasking thresholds started at $25\sigma$ to produce clean initial models, then decreased iteratively to $3\sigma$ in the final imaging cycle for optimal recovery of faint structures. Self-calibration solutions were derived using CubiCal \citep{Kenyon2018} tasks integrated within CARACal.

For the subtraction of discrete-point-sources in MeerKAT, we used the same method as for uGMRT. First, to get the model for the point sources, we image the data using uv baselines greater than $5\rm k\lambda$ and Briggs weighting of $-2.0$. Then we put that model in the model column of the final self-calibrated data. After subtracting the model from the original visibilities, we image the residual visibilities using uv baselines $0.2-10\rm k\lambda$, a Gaussian taper of 10$''$, and Briggs weighting of zero to get the image of the diffuse radio emission. All the final images, both from uGMRT and MeerKAT, are smoothed to a beam of $20''\times20''$.
 
\begin{figure*}
    \centering
    \includegraphics[width = 1\textwidth]{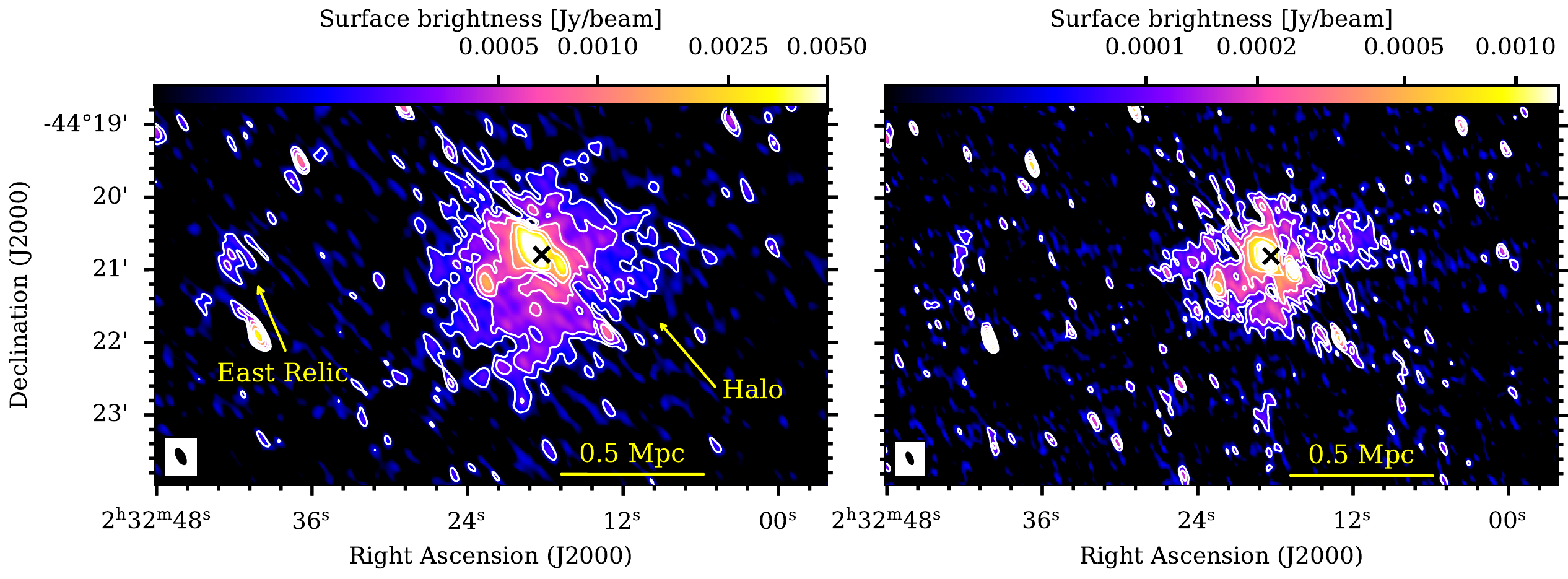}
    \caption{Left: the uGMRT 400 MHz full-resolution image of the cluster RXCJ0232 is show in colour and contours. The beam size of the image is 14.7$^{\prime\prime}\times$ 6.6$^{\prime\prime}$ with a position angle (PA) of 27.83$^{\circ}$. The contour levels are [1, 2, 4, 8, \dots] × 3$\sigma_{rms}$ with $\sigma_{rms}$ = 43 $\mu$Jy beam$^{-1}$. The positions of the halo and the candidate east relic are marked with yellow arrows. Right: the uGMRT 650 MHz full-resolution image with a beam size of $10.8^{\prime\prime}\times4.6^{\prime\prime}$ (PA $=23.65^{\circ}$) is shown. The contour levels are [1, 2, 4, 8, \dots] × 3$\sigma_{rms}$ with $\sigma_{rms}$ = 20 $\mu$Jy beam$^{-1}$. The position of the central radio-loud BCG (BCG$-$A) is marked with a black cross.}
    \label{fig: full}
\end{figure*}

\section{Results} \label{sec: result}

\subsection{Continuum Images}

\begin{figure*}
    \centering
    \includegraphics[width = 1\textwidth]{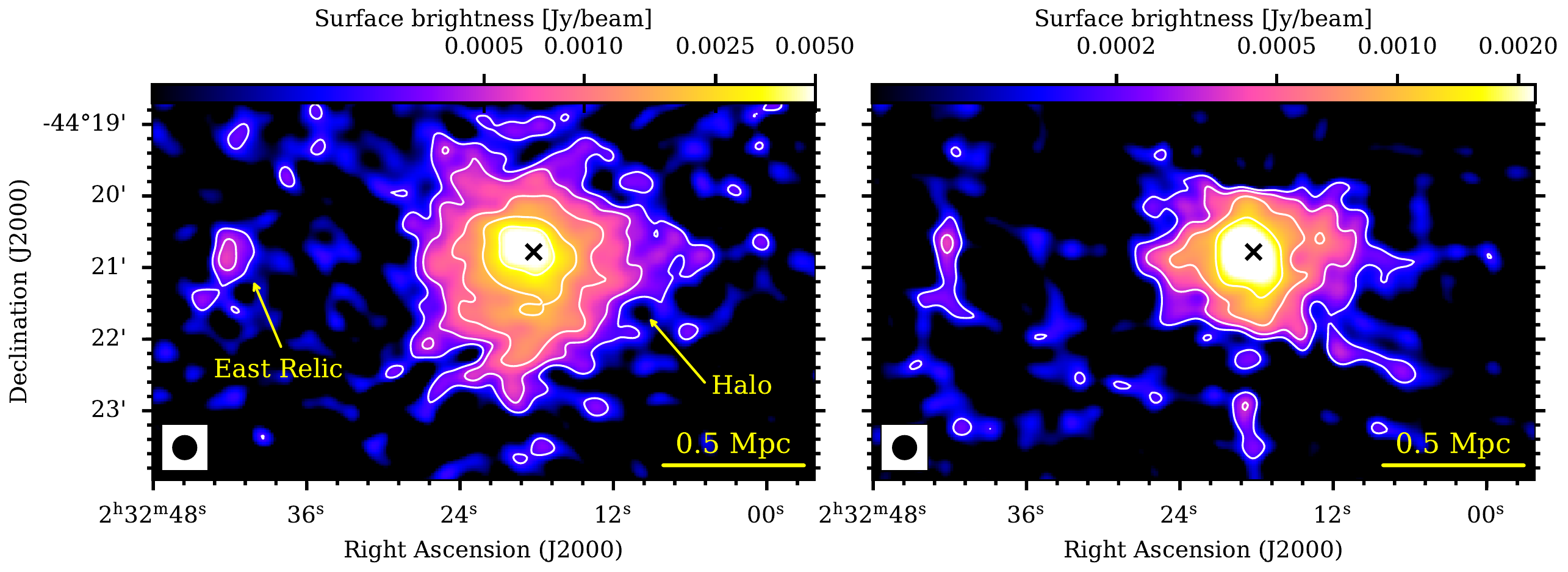}
    \includegraphics[width = 0.5\textwidth]{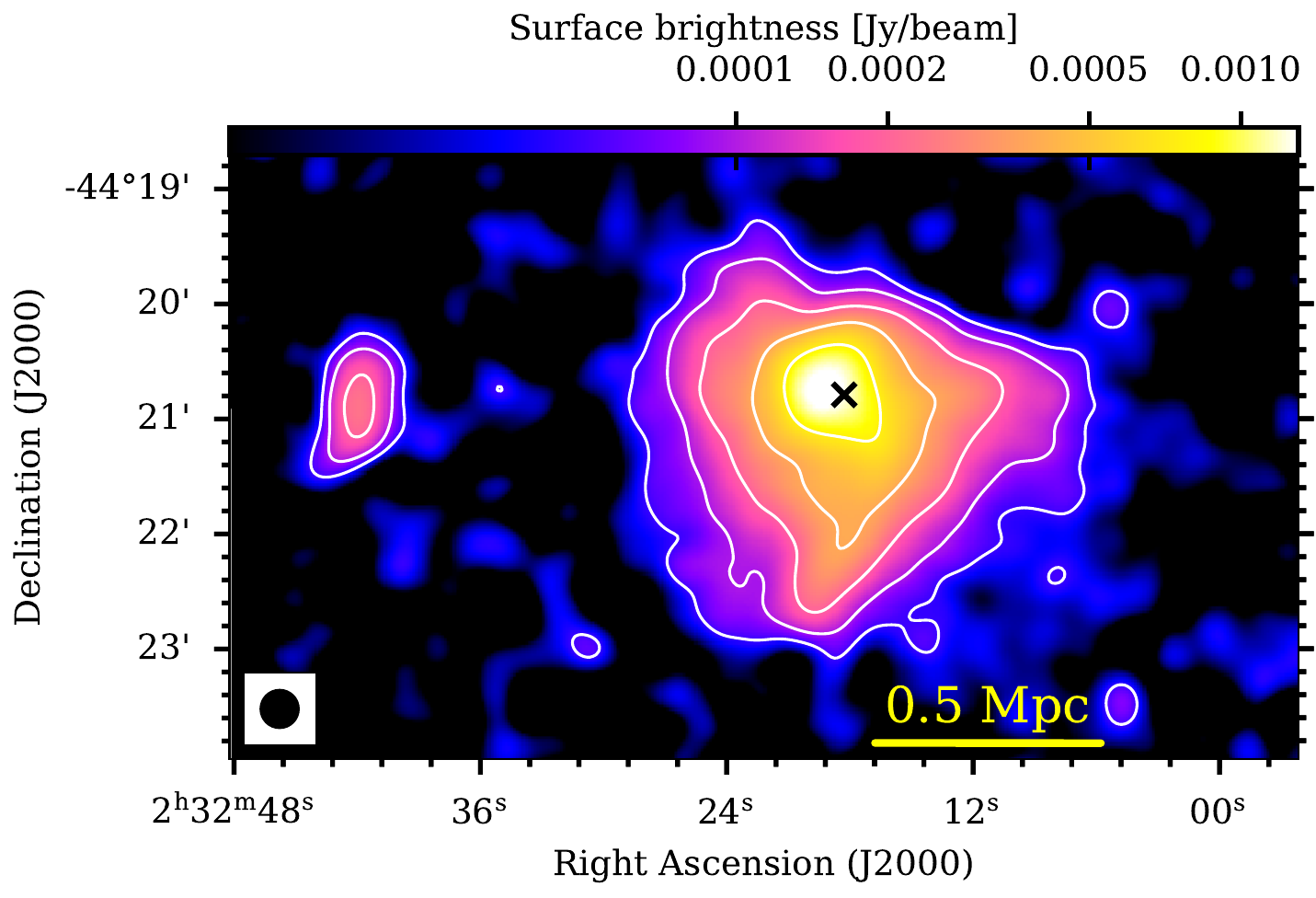}
    \caption{\textit{Top$-$left}: uGMRT 400 MHz point-source-subtracted low-resolution (20$^{\prime\prime}\times$ 20$^{\prime\prime}$, PA $=0^{\circ}$) image of the central region of cluster RXCJ0232 is shown in colour and contours. The contour levels are [1, 2, 4, 8, \dots] × 3$\sigma_{\rm rms}$ with $\sigma_{\rm rms}$ = 79 $\mu$Jy beam$^{-1}$. The central radio-loud BCG (BCG$-$A) is marked with a black cross. The positions of the halo and the candidate east relic are also marked with yellow arrows. \textit{Top$-$right}: uGMRT 650 MHz low-resolution image of the cluster shown in contours and colour. The contour levels are the same as the left panel with $\sigma_{\rm rms}$ = 56 $\mu$Jy beam$^{-1}$. \textit{Bottom}: MeerKAT 1283 MHz low-resolution point-source-subtracted image of the cluster with similar colours and contour levels with $\sigma_{\rm rms}$ = 16 $\mu$Jy beam$^{-1}$.}
    \label{fig: diffuse}
\end{figure*}

Figure \ref{fig: rgb} presents a multiwavelength view of RXCJ0232, combining optical, X-ray, and radio observations. The optical r-band emission (yellow) from the Dark Energy Spectroscopic Instrument (DESI) Legacy Imaging Survey \citep{Dey2019} traces the distribution of galaxies within the cluster. The diffuse X-ray emission (blue) detected by Chandra reveals the hot intracluster medium \citep{Parekh2021}, indicating the gravitational potential well of the cluster. The MeerKAT 1283 MHz radio continuum emission (red) highlights synchrotron-emitting sources \citep{Knowles2022}, including the candidate east relic and radio halo associated with the cluster, as well as discrete radio galaxies within and behind the cluster. Figure \ref{fig: full} shows the central portion of the high-resolution uGMRT band$-3$ and $-4$ images of the cluster RXCJ0232. The source marked by the cross corresponds to one of the central BCGs \citep[BCG$-$A;][]{Kale2019}. The other BCG (BCG$-$B) does not have any radio counterpart. We can see diffuse radio emission from the ICM surrounding the central BCG, along with several radio sources. The uGMRT band$-$3, $-4$ and MeerKAT low-resolution ($20^{\prime\prime}\times20^{\prime\prime}$) point-source-subtracted images of the diffuse emissions are presented in Figure \ref{fig: diffuse}. The largest extent of the halo captured in both bands 3 and 4 is $\sim 1.1$ Mpc (249$^{\prime\prime}$), and that for the MeerKAT image is $\sim 1.2$ Mpc (278$^{\prime\prime}$). The candidate east relic has been detected at both uGMRT frequencies \citep{Kale2022}. The largest linear extents of the candidate east relic at 400 MHz and 650 MHz are 305 kpc (72$^{\prime\prime}$) and 297 kpc (70$^{\prime\prime}$), respectively. In the MeerKAT image, the largest extents of the candidate east and south relics are 361 kpc (85$^{\prime\prime}$) and 552 kpc (130$^{\prime\prime}$), respectively. We did not detect the south relic \citep{Kale2022} in any uGMRT band, likely due to contamination from a nearby bright radio source and lower sensitivity than MeerKAT. We have scaled the $3\sigma_{\mathrm{rms}}$ flux density ($48 \rm \ \mu Jy \ beam^{-1}$) of the 1283 MHz image to 400 MHz with a spectral index of $-1$, which is $\sim 1.5$ times lower than the $3\sigma_{rms}$ flux density of the 400 MHz image.

\begin{table*}
	\centering
	\caption{Various properties of the halo and the candidate east relic}
	\begin{tabular}{@{}lcccccc@{}}
		\hline\hline
		Type of   & Telescope \& & Flux           & Size               & Integrated         & Equipartition    & Radio Power       \\
        Radio     & Frequency    & Density        &                    & Spectral           & Magnetic Field   & P$_{\rm 1.4 GHz}$      \\
        Source    & (MHz)        & (mJy)          & (kpc $\times$ kpc) & Index              &  ($\mu$G)        & (W Hz$^{-1}$)       \\
        \hline\hline
		           & uGMRT 400    & 82.9 $\pm$ 9.3 & 1058 $\times$ 862  &                    &                  &                       \\
		Halo      & uGMRT 650    & 37.5 $\pm$ 4.2 & 1058 $\times$ 637  & $-1.17 \pm 0.17$   & $3.86 \pm  0.31$ & $(4.98 \pm 1.06) \times 10^{24}$  \\
		           & LGMRT 606    & 52.0 $\pm$ 5.6 & 892 $\times$ 807   &                    &                  &                         \\
                  & MeerKAT 1283 & 21.0 $\pm$ 2.3 & 1181 $\times$ 1062 &                    &                  &                          \\
        \hline
        Candidate & uGMRT 400    & 2.0 $\pm$ 0.3  & 305 $\times$ 174   &                    &                  &                            \\
		  East      & uGMRT 650    & 1.0 $\pm$ 0.2  & 297 $\times$ 149   & $-0.85 \pm 0.17$   & $1.55 \pm 0.15$  & $(1.66 \pm 0.48) \times 10^{23}$         \\
		Relic     & LGMRT 606    & 1.4 $\pm$ 0.3  & 233 $\times$ 140   &                    &                  &                              \\
                  & MeerKAT 1283 & 0.7 $\pm$ 0.1  & 361 $\times$ 204   &                    &                  &                               \\
		\hline
	\end{tabular}
        \label{table2}
    \tablecomments{The sizes reported here are measured from the low-resolution ($20^{\prime\prime}$) images.}
\end{table*}

\subsection{Integrated Flux Density and Spectral Index}

The total flux densities of the radio halo and the candidate east relic are calculated from the discrete point-source subtracted low-resolution (20$^{\prime\prime}$) images. The same areas have been selected for halo and relic to determine the total flux densities at all frequencies. The uncertainties in the flux densities are measured by the following formula:

\begin{eqnarray}
	\Delta \rm S = \sqrt{(\sigma_{\rm abs} \rm S)^2 + (\sigma_{\rm rms}\sqrt{\rm N_b})^2 + (\sigma_{sub}S)^2}
	\label{Eq1}
\end{eqnarray}

\begin{figure}
    \centering
    \includegraphics[width = 0.49\textwidth]{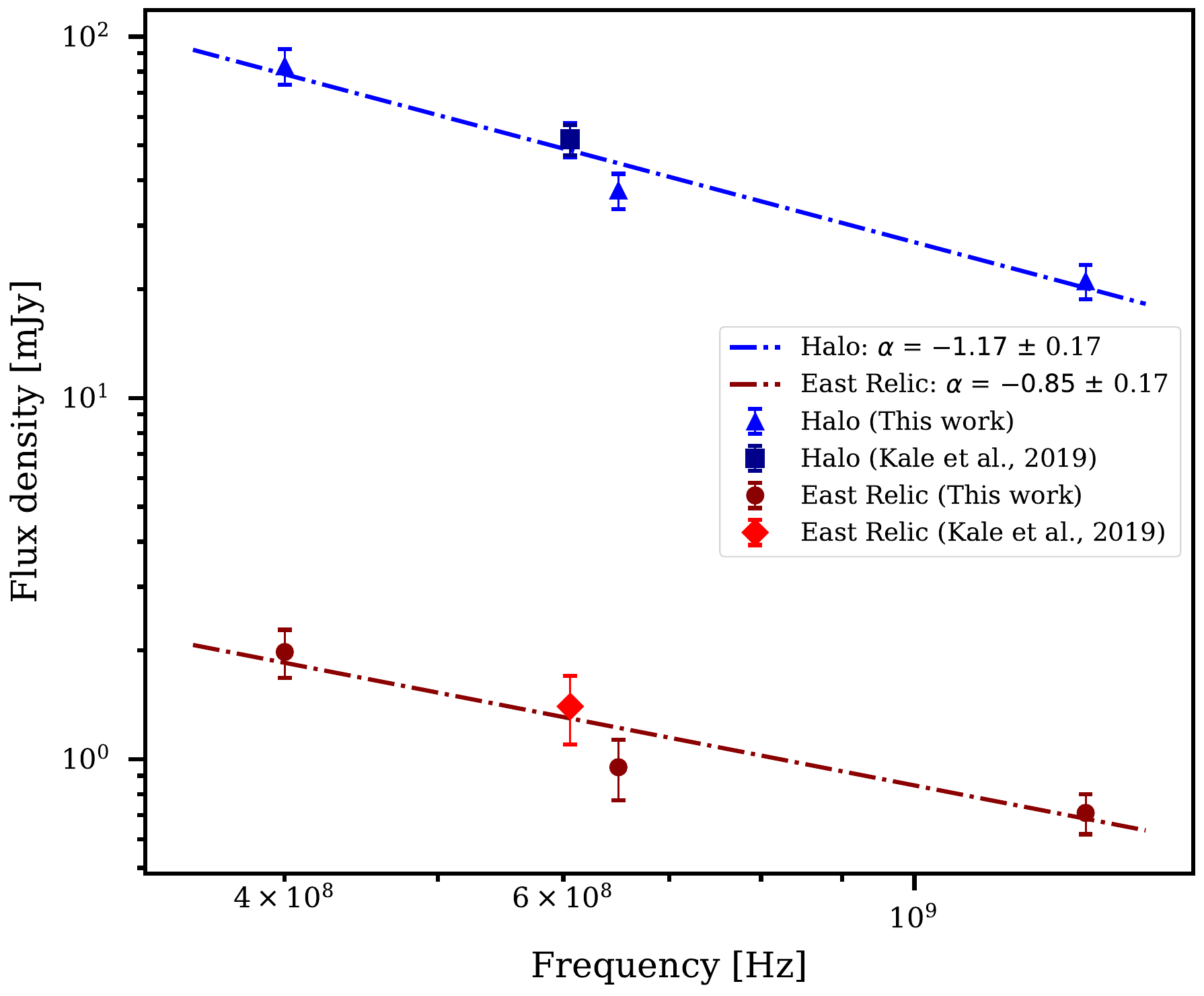}
    \caption{The integrated spectrum of the halo and the candidate east relic, where the radio halo measurements from uGMRT at 400 and 650 MHz and MeerKAT at 1283 MHz are shown as blue triangles, with the 606 MHz point from \citet{Kale2019} shown as a dark blue square. The best-fit power law is shown as a blue dotted line. Flux densities of the candidate east relic at 400, 650, 1283, and 606 MHz are shown as dark red circles and a red diamond, with a best-fit power law as a dark red dotted line.}
    \label{fig: intx}
\end{figure}

Where $\sigma_{\mathrm{abs}}$ is the calibration error in the absolute flux density, N$_\mathrm{b}$ is the number of beams in the selected region, $\sigma_{\mathrm{rms}}$ is the rms noise in the image, $\sigma_{\rm sub}$ is the point source subtraction error, and S is the flux density of the radio emission. We have considered $\sigma_{\mathrm{abs}}$ of 10\% for all the frequencies \citep{Chandra2017,Balboni2026}. The various properties of the halo and the relics are tabulated in Table \ref{table2}, including the flux densities and sizes. In Figure \ref{fig: intx}, we plot flux density versus frequency on logarithmic axes, including the LGMRT 606 MHz point from \citet{Kale2019}. Using the relation:

\begin{eqnarray}
	\log(\rm S_{\nu}) = \alpha\log(\nu)+\rm c
	\label{Eq2}
\end{eqnarray}

where $\rm S_\nu$ is the flux density at frequency $\nu$, and $\alpha$ is the spectral index. We fit a straight line to derive the spectral index $\alpha$. The halo exhibits a spectral index of $\alpha = -1.17 \pm 0.17$, while the candidate east relic shows a value of $\alpha = -0.85 \pm 0.17$ (Figure \ref{fig: intx}). Using these values, we estimate the extrapolated 1.4 GHz flux densities $19.1 \pm 4.0$ mJy for the halo, and $0.7 \pm 0.2$ mJy for the candidate east relic. The 1.4 GHz radio power ($K$-corrected) can be calculated from the relation.

\begin{eqnarray}
    \rm P_{\rm 1.4 \ GHz}=\frac{4\pi \rm D_L^2}{(1+\rm z)^{(\alpha + 1)} } \times \rm S_{\rm 1.4 \ GHz}
    \label{Eq3}
\end{eqnarray}

where D$_{\mathrm{L}}$ is the luminosity distance to the cluster. The 1.4 GHz radio powers for the halo and the relic are $(4.98 \pm 1.06) \times 10^{24}$ W Hz$^{-1}$ and $(1.66 \pm 0.48) \times 10^{23}$ W Hz$^{-1}$, respectively.

Assuming equipartition between relativistic particles and the magnetic field \citep{Govoni2004}, we estimated the magnetic field strengths for the diffuse sources. Taking a ratio of relativistic proton to electron energy of $k = 1$ and Lorentz factor $\gamma_{\rm min} = 100$, we derived magnetic field values of $3.86 \pm 0.31 \ \mu\text{G}$ for the giant radio halo and $1.55 \pm 0.15 \ \mu\text{G}$ for the candidate east relic.

\begin{figure}
    \centering
    \includegraphics[width = 0.49\textwidth]{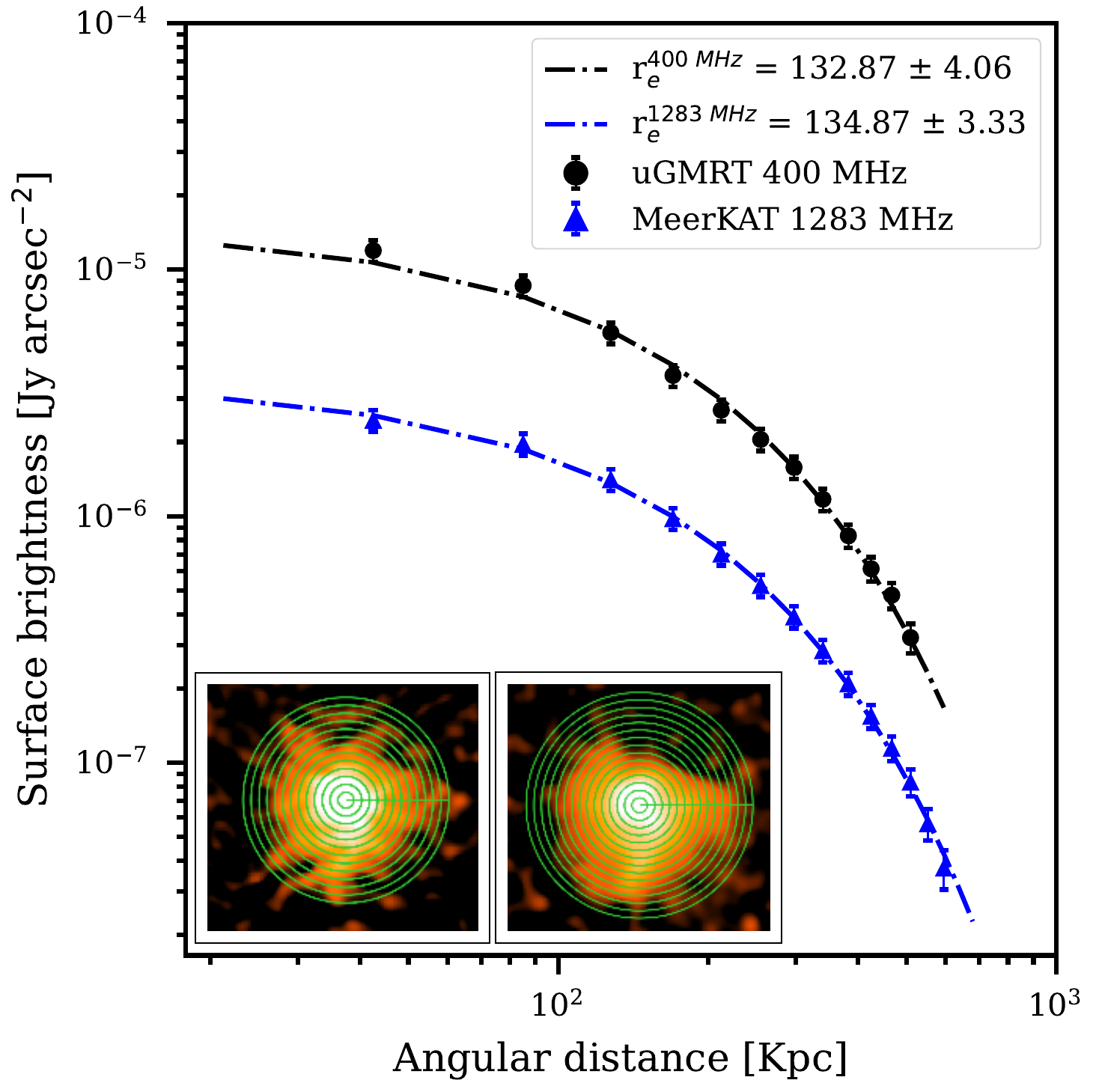}
    \caption{Radio surface brightness profiles of the halo at uGMRT 400 MHz (black dotted line) and MeerKAT 1283 MHz (blue dotted line). The annotated images (400 MHz on the left and 1283 MHz on the right) show the radio maps with the annular regions overlaid. The surface brightness values are estimated within each annular region, and the uncertainties in the profiles are estimated considering the measurement errors.}
    \label{fig: profile}
\end{figure}

\begin{table}
  \centering
  \caption{Best fit parameters of the Radio surface brightness profile of the halo.}
  \begin{tabular}{@{}ccc@{}}
    \hline
      Frequency (MHz) & I$_{0}$ & r$_{\rm e}$  \\
    \hline\hline

    400 & 14.72 $\pm$ 0.98 & 132.87 $\pm$ 4.06  \\

    1283 & 3.51$\pm$ 0.21 & 134.87$\pm$ 3.33   \\
    
    \hline
  \end{tabular}
    \label{table3}
  \tablecomments{Column (1): frequency used for the analysis. Column (2): central brightness of the fit in units of $\mu$Jy \ arcsec$^{-2}$. Columns (3): e-folding radius in kpc.}
 \end{table}

\subsection{Radio Surface Brightness Profile}

Following \citet{Murgia2009}, we modelled the radial variation of radio surface brightness for the halo, using a simple exponential law:

\begin{eqnarray}
    \rm I(r) = I_0 e^{-\frac{r}{r_e}}
    \label{Eq9}
\end{eqnarray}

where I$_0$ is the central brightness, $\rm I(r)$ is the brightness at a distance $\rm r$ of the halo, and $\rm r_{\rm e}$ is the e-folding radius, defined as the radius where the surface brightness drops to $\rm I_0/e$.

For uGMRT 400 MHz and MeerKAT 1283 MHz data, we modelled the radio surface brightness using the same procedure. We did not fit the surface brightness profile for the uGMRT 650 MHz for the reason stated in Appendix \ref{issue}. The radio-emitting region was divided into concentric annuli centred on BCG-A. The diameter of the central annulus was set to the synthesised beam size ($20^{\prime\prime}$), and the radii of subsequent annuli were increased in steps of half a beam. We measured the total radio brightness in each annulus and fitted the values using equation \ref{Eq9}. Figure \ref{fig: profile} shows the resulting surface brightness profiles at both frequencies on a log–log scale, with black and blue points indicating the measurements and lines showing the best-fit models. The best-fitting parameters are listed in Table \ref{table3}. Considering the fitting uncertainty, we do not see any frequency evolution of the e-folding radius.

\begin{figure*}
    \centering
    \includegraphics[width = 1\textwidth]{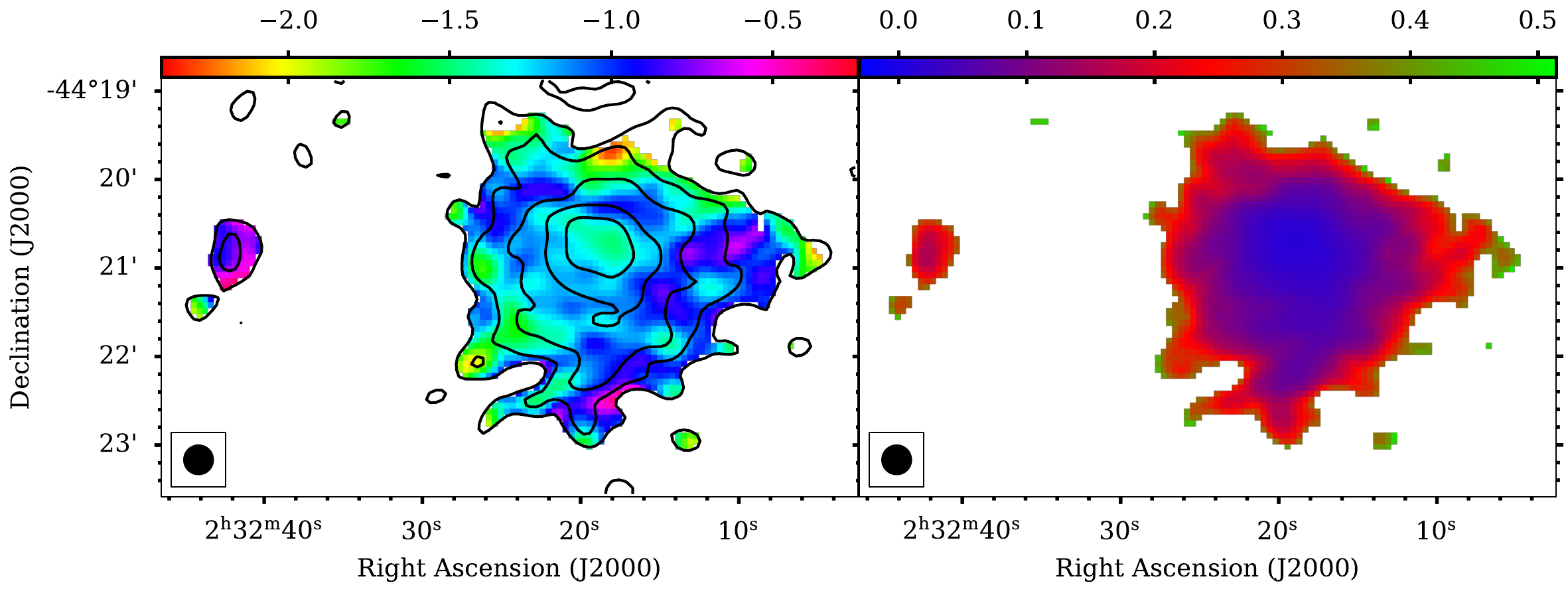}
    \caption{The resolved spectral index map (\textit{left}) at $20^{\prime\prime}$, between the uGMRT 400 MHz and MeerKAT 1283 MHz images, of the central region of cluster RXCJ0232, with the corresponding error map (\textit{right}), is shown in colour. The contour levels in black are [1, 2, 4, 8, \dots] $\times 3\sigma_{\rm rms}$ ($\sigma_{\rm rms}=79$ $\mu$Jy beam$^{-1}$) from the 400 MHz image of uGMRT. See appendix \ref{rsindxGMRT} Figure \ref{fig: rindxgmrt} for the resolved spectral index map between the uGMRT 400 and 650 MHz images.}
    \label{fig: rindx}
\end{figure*}

\subsection{Resolved Spectral Analysis}
 
A spatially resolved spectral index map offers a deeper understanding of the underlying acceleration mechanism of charged particles at local scales \citep[e.g.,][]{Brunetti2014, Weeren2016, Rajpurohit2023, santra2026}. We produced two spectral index maps: one using the uGMRT 400 and 650 MHz images, and another combining uGMRT 400 MHz with MeerKAT 1283 MHz. Point-source-subtracted, low-resolution images were made, restricted to the same baseline range (0.2–10 k$\lambda$), and then smoothed to a common $20^{\prime\prime}$ beam using the \texttt{imsmooth} task in \texttt{CASA}. This ensures similar resolution across all bands and consistent detection of the halo. Pixels below 2.5 $\sigma_{\mathrm{rms}}$ were masked to reduce uncertainties. Flux densities were resampled 1000 times, assuming a Gaussian distribution with mean equal to the measured flux and standard deviation equal to the image rms. The spectral index and its uncertainty were then computed pixel-wise as the mean and standard deviation of the 1000 realisations\footnote{\url{https://github.com/revoltek/scripts/blob/master/spidxmap.py}} \citep[see][for details]{deGasperin2017}.

The resolved spectral index alongside the corresponding error map for the uGMRT 400 MHz and MeerKAT 1283 MHz is presented in Figure \ref{fig: rindx}. The same for the uGMRT 400 and 650 MHz is shown in appendix \ref{rsindxGMRT} Figure \ref{fig: rindxgmrt}. Despite the problem (see Appendix \ref{issue}) present in the uGMRT band 4 visibilities, the spectral index map of the cluster core region is reasonably accurate. The candidate east relic is very small, making comparisons of spectral indices across different regions difficult. Also, the error values are high in the relic region. Therefore, we cannot comment on the resolved spectral index of the relic. The error map indicates that typical error values are less than 0.3, except for the edges over the halo region. The spectral index maps show some local fluctuations, with values across the radio halo mostly ranging between $-1.0$ and $-1.3$. Even with these fluctuations on local scales, the emission globally lies predominantly within this range, indicating a relatively less-steep spectrum with little local variation. This suggests turbulence is occurring on small scales throughout the cluster region.

\begin{figure*}
    \centering
    \includegraphics[width = 0.495\textwidth]{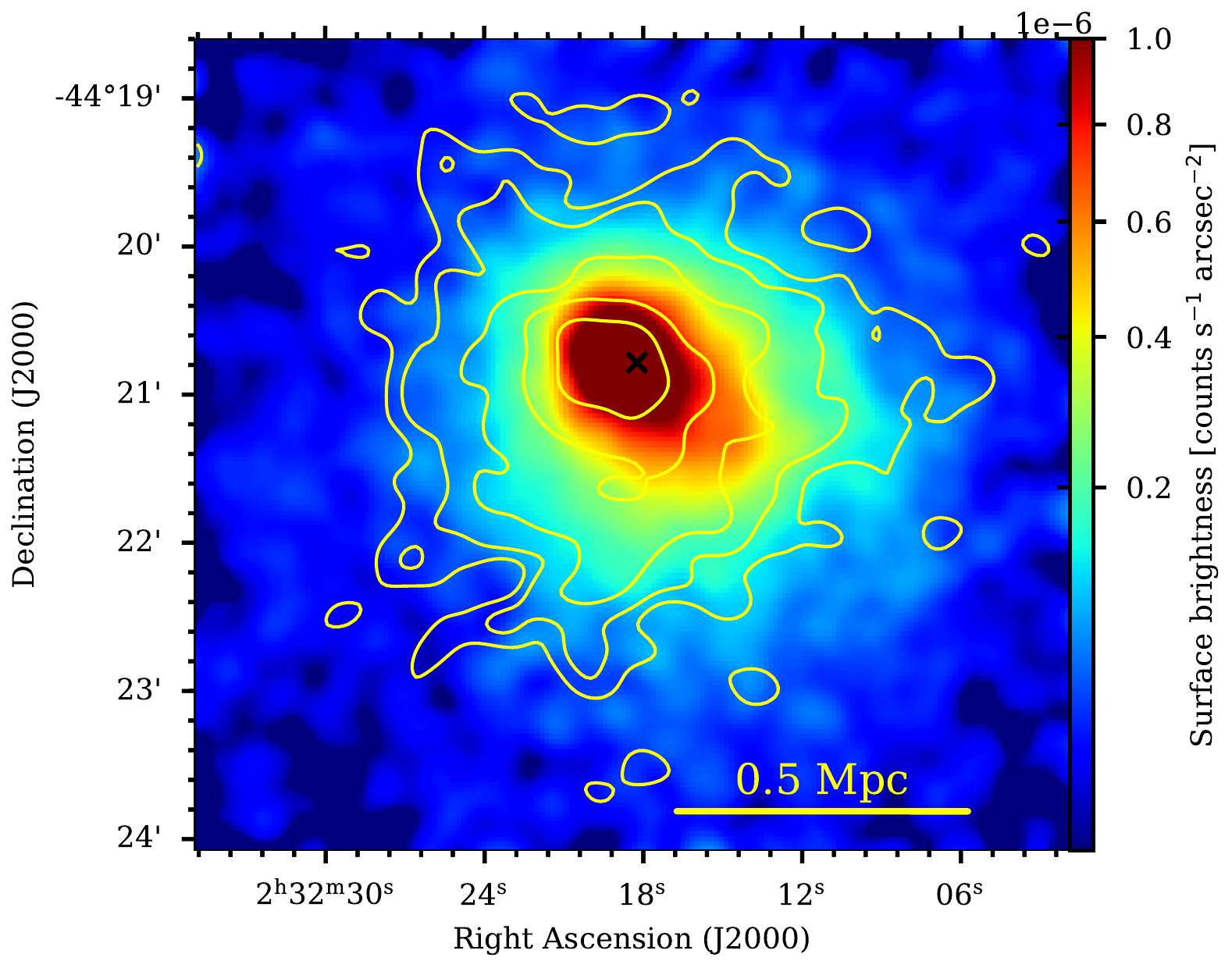}
    \includegraphics[width = 0.495\textwidth]{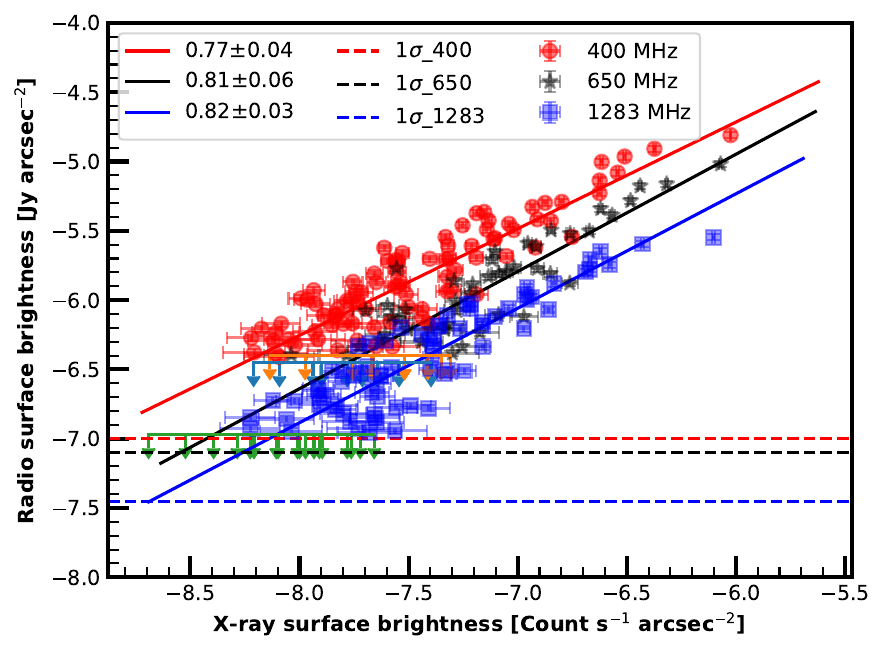}
    \caption{\textit{Left:} Chandra X-ray image (background-subtracted and exposure-corrected) of RXCJ0232 in the 0.5–7 keV band, smoothed with a 3 pixel Gaussian. Overlaid are low-resolution ($20^{\prime\prime}$) uGMRT 400 MHz radio contours (yellow) at [1, 2, 4, 8, \dots] × 3$\sigma_{\rm rms}$, with $\sigma_{\rm rms}$ = 79 $\mu$Jy beam$^{-1}$. \textit{Right:} radio vs X-ray surface brightness relation for RXCJ0232. Data points above 3$\sigma_{\rm rms}$ at 400, 650, and 1283 MHz are shown as red circles, black stars, and blue squares, respectively. Arrows denote 2$\sigma_{\rm rms}$ upper limits, while dashed lines indicate the 1$\sigma_{\rm rms}$ levels at each frequency. Best-fit point-to-point correlations are overlaid.}
    \label{fig: xray}
\end{figure*}

\subsection{Radio and X-Ray Correlation}

Figure \ref{fig: xray} (left panel) shows the Chandra exposure-corrected and background-subtracted X-ray map (taken from \citealt{Parekh2021}) of the cluster RXCJ0232 with 400 MHz uGMRT radio contours overlaid. The overall morphologies of the X-ray and radio emission from the cluster are very similar; however, in the southwest, the X-ray emission extends farther than the radio emission. The positions of radio and X-ray emission peaks are the same as those of the central radio-loud BCG (BCG-A).

Point-to-point correlation between X-ray and radio surface brightnesses offers insight into how relativistic particles and magnetic fields spatially relate to the X–ray–emitting thermal gas \citep[e.g.,][]{Govoni2001}. This method is motivated by the striking morphological similarity between radio halos and X-ray emission, which suggests an energetic link between the nonthermal plasma and the thermal ICM, and has been carried out on many systems \citep[e.g.,][]{Giacintucci2005, Hoang2019, Cova2019, Xie2020, Rajpurohit2021a, Rajpurohit2021b, Bruno2021, Bonafede2022, Riseley2022a, Santra2024,balboni24, riseley24, santra24b, kale25, santra2026, vanweeren2026}. Moreover, such analyses can help discriminate between different mechanisms powering relativistic electrons, such as hadronic collisions and/or merger-driven turbulent reacceleration \citep{Govoni2001, Ignesti2020}.

Sensitive radio and X-ray maps allow a detailed spatial comparison of the nonthermal and thermal ICM using the Point-to-point TRend EXtractor \citep[\texttt{PT-REX}\footnote{\url{https://github.com/AIgnesti/PT-REX}};][]{Ignesti2020, Ignesti2022}. We first masked all regions outside the radio halo (regions with flux density $<2\sigma_{\rm rms}$ and not part of the halo emission) and then created a $20^{\prime\prime}$ ($\sim 85$ kpc) square grid across the halo. For each cell, we measured the mean and rms of the radio and X-ray surface brightnesses. The resulting data points were fitted with the standard power-law relation commonly adopted in the literature, given by:

\begin{eqnarray}
	\log(\rm I_{\rm R}) = \rm a + b \log(\rm I_{\rm X})
	\label{Eq10}
\end{eqnarray}

Here, $\rm I_{\rm R}$ and $\rm I_{\rm X}$ denote the radio and X-ray surface brightnesses. The slope b characterises the radio–X-ray correlation, where $\rm b=1$ indicates a linear relation, $\rm b<1$ a sublinear trend, and $\rm b>1$ a superlinear one. In general, leptonic models predict sublinear correlations, while hadronic models favour superlinear slopes. For the fit, we used only grid cells with radio emission above $3\sigma_{\rm rms}$; cells between $2\sigma_{\rm rms}$ and $3\sigma_{\rm rms}$ were treated as upper limits. To account for intrinsic scatter and incorporate the upper limits, we used \texttt{LinMix}\footnote{\url{https://linmix.readthedocs.io/en/latest/src/linmix.html}}
 \citep{Kelly2007}, a Markov Chain Monte Carlo-based hierarchical regression.

There is a positive correlation (Figure \ref{fig: xray}) between the radio and X-ray surface brightnesses at all frequencies, with a correlation coefficient of $0.85$ and a correlation slope of $\sim 0.80$. We find a sublinear slopes of b$_{400 \rm MHz}$ = 0.77 $\pm$ 0.04, b$_{650 \rm MHz}$ = 0.81 $\pm$ 0.06, and b$_{1283 \rm MHz}$ = 0.82 $\pm$ 0.03, with an intrinsic scatter of $\sim$ 0.06 at all frequencies (See Table \ref{corr_res}). This scatter likely arises from both measurement uncertainties and intrinsic dispersion due to the physical properties of the halo, such as inhomogeneous turbulence and fluctuations in the magnetic field and the density of relativistic electrons. The areas chosen for the I$_{\rm R}$--I$_{\rm X}$ correlation analysis vary slightly with frequency because of the signal-to-noise variations at the edge of the halo and the different sizes of the halo at different frequencies. We repeated the fitting procedure, restricting the analysis to regions where emission exceeds the $3\sigma$ threshold for all the frequencies. This approach resulted in slope values of $\rm b_{400\,\mathrm{MHz}} = 0.76 \pm 0.05$, $\rm b_{650\,\mathrm{MHz}} = 0.80 \pm 0.06$, and $\rm b_{1283\,\mathrm{MHz}} = 0.81 \pm 0.04$. These slopes matched well with the previous ones, and the general trend also remains the same (sublinear). We also do not find a frequency evolution of the correlation slopes, similar to other giant radio halos \citep[e.g.,][]{Hoang2019, Rajpurohit2021b, Rajpurohit2023, santra24b}. 

\begin{table}
  \centering
  \caption{Linmix fitting slopes and correlation coefficients for Figure~\ref{fig: xray}.}
  \begin{tabular}{@{}ccccc@{}}
    \hline
      Frequency ($\nu)$ & Slope (b) & r$_{\rm s}$ & r$_{\rm p}$  \\
    \hline\hline

    400 & 0.77 $\pm$ 0.04 & 0.86 & 0.87 \\

    650 & 0.81 $\pm$ 0.06 & 0.83 & 0.87  \\

    1283 & 0.82 $\pm$ 0.03 & 0.86 & 0.89  \\

    \hline
  \end{tabular}
  \label{corr_res}
  \tablecomments{Column (1): frequency used for the analysis in MHz. Column (2): correlation slope, including the upper limits. Column (3): Spearman coefficient. Column (4): Pearson coefficient}
\end{table}

\section{Discussion} \label{sec: discussion}

RXCJ0232 exhibits a combination of properties that place it at the interface between classical minihalo and giant radio halo systems. The diffuse radio emission extends to a scale of $\sim1.2$ Mpc, comparable to that of giant radio halos, while remaining centred on the brightest cluster galaxy and closely coincident with the peak of the X-ray emission, as typically observed in relaxed clusters hosting minihalos. At the same time, the \textit{Chandra} X-ray surface brightness distribution appears largely regular, with no clear signatures of a major merger, aside from a broad extension toward the southwest. In the following, we discuss the dynamical, spectral, and morphological properties of RXCJ0232 in the context of particle acceleration mechanisms and the origin of its diffuse radio emission.

\subsection{Cluster Dynamics and Radio Emission}

The X-ray morphology of RXCJ0232 broadly resembles that of a relaxed cluster, with a centrally peaked surface brightness distribution and no clear signatures of a major merger (Figure~\ref{fig: xray}). Apart from an overall elongation in the northeast to southwest direction, the X-ray image shows no strong disturbances, such as multiple peaks, sharp discontinuities, or large centroid shifts. This makes RXCJ0232 particularly intriguing, as it hosts large-scale diffuse radio emission despite exhibiting several characteristics typical of relaxed, cool-core systems. Accordingly, its dynamical state has been classified inconsistently in the literature. Based on power ratios ($P_3/P_0$) and centroid shifts, \citet{Weissmann2013} classified the cluster as moderately disturbed, whereas \citet{Lovisari2017}, using a broader set of morphological parameters, identified it as relaxed. Consistent with this ambiguity, \citet{Parekh2021} showed that RXCJ0232 does not fall cleanly into either relaxed or disturbed category when quantified using Gini, $M_{20}$ \citep{Lotz2004, Lotz2008}, and concentration parameters, indicating an intermediate dynamical state.

Thermodynamically, RXCJ0232 shows clear evidence for a cool core. Its central cooling time of $\sim 1$ Gyr \citep{Cavagnolo2009} satisfies the criterion for strong cool-core clusters \citep{Santos2008}. However, its central entropy of $\sim 50\ \rm keV\ cm^2$ places it in the weak cool-core regime according to the classification of \citet{Hudson2010}, again suggesting an intermediate system. A clear substructure is detected $\sim1^{\prime}$ southwest of the cluster centre, aligned with the large-scale X-ray elongation \citep{Parekh2021}. This feature is visible in both unsharp-masked and $\beta$-model–subtracted images and has a characteristic size of $\sim0.35^{\prime}$. The presence of this substructure, despite the largely undisturbed core, points to a minor merger or ongoing accretion event, likely in a preferred filamentary direction. However, the presence of the two BCGs $\sim 100$ kpc apart in the cluster centre \citep{Pierini2008} supports the minor merger scenario.  See Table \ref{tab:dynamics_summary} where we summarise the ambiguous dynamical state reported in the literature about RXCJ0232, including morphological parameters and thermodynamic indicators. The coexistence of a cool core and a giant radio halo in RXCJ0232 therefore suggests a scenario in which the cluster core remains intact, while turbulence generated by mild disturbances at larger radii is sufficient to power the extended radio emission. 

\begin{table*}
\caption{Summary of dynamic and thermodynamic indicators for RXCJ0232-4420.}
\label{tab:dynamics_summary}
\centering
\begin{tabular}{l l l l}
\hline\hline
\textbf{Parameter} & \textbf{Value/Status} & \textbf{Classification/Comment} & \textbf{Reference} \\
\hline\hline
-- & Presence of two BCGs & Evidence of merger & \citet{Pierini2008}\\
\hline
Central entropy ($K_0$) & $44.62 \pm 12.42$ keV cm$^2$ & Intermediate between mini and radio halos & \citet{Giacintucci2017} \\
 &  & ``weak'' cool-core & \citet{Hudson2010} \\
\hline
Centroid shift ($w$) & $w\times10^{-2}=1.97\pm0.04$ & &\\
Concentration ($c$) & $c=0.31\pm0.01$ & Relaxed system & \citet{Lovisari2017} \\
\hline
Central temperature & $5.71\pm0.77$ keV & & \\
Central cooling time & $1.38 \pm 0.19$ Gyr & ``weak'' cool-core & \citet{Cavagnolo2009} \\
\hline
Power ratio ($P_3/P_0$) & $(1.45\pm0.55)\times10^{-7}$ &  &  \\
Centroid shift ($w$) &  $(2.09\pm0.06)\times10^{-2}$ & Moderately disturbed/complex & \citet{Weissmann2013}\\
\hline
Gini & $0.55\pm0.005$ & & \\
$M_{20}$ & $-1.58\pm0.046$ & &\\
Concentration & $1.37\pm0.014$ & Intermediate dynamical state & \citet{Parekh2021} \\
\hline
-- & No shocks/cold fronts detected & Core remains undisturbed & \citet{Parekh2021} \\
\hline
Substructure & Detected $\sim1'$ southwest of core & Evidence of minor merger/accretion & \citet{Parekh2021} \\
\hline\hline
\end{tabular}
\end{table*}

\subsection{Possible Origin of the Radio Halo}

Giant radio halos in merging clusters follow well-established correlations between radio power and cluster mass or mass proxies, commonly interpreted as a consequence of merger-driven turbulence that amplifies magnetic fields and re-accelerates relativistic particles \citep[e.g.,][]{Cassano2010, Cassano2012, basu12, kale15}. While minor mergers can also generate radio halos, such systems are rare and show steeper radio spectra due to lower available turbulent energy \citep[e.g.,][]{Cassano2006, Brunetti2008}. In the turbulent reacceleration framework, relaxed cool-core clusters are expected to host giant radio halos with steeper spectra, as the turbulent energy available outside the core is lower than in major mergers. The apparent rarity of such systems is likely an observational bias, as the steep-spectrum emission is difficult to detect with standard surveys, but becomes visible at low frequencies or with high-sensitivity instruments like LOFAR, MeerKAT and uGMRT. This expectation is consistent with low-frequency discoveries, which preferentially detect such halos in less disturbed systems due to their steep spectra \citep[e.g.,][]{cassano23}. The integrated spectral index of the RXCJ0232 halo ($\alpha = -1.17 \pm 0.17$) is comparable to the average values reported for giant radio halos in massive merging clusters, which typically range between $-1.2$ and $-1.3$. In the context of the turbulent reacceleration, this value is relatively less steep than what might be expected for a system with a cool core, where lower turbulent energy is typically available. This suggests that recent or efficient injection of turbulence may be occurring despite the intermediate dynamical state of the cluster \citep{Kale2019, Parekh2021}.

\citet{Kale2019} had reported that the emission of this cluster falls well within the correlation (radio power vs mass) of giant radio halos. Despite the improved sensitivity of MeerKAT at 1.28 GHz, we still found that the cluster lies near the expected correlation for giant halos. The central surface brightness (I$_{0}$) and characteristic radius ($r_{\mathrm{e}} \sim 130$ kpc) are also consistent with those of known giant radio halos, lying within the observed dispersion of the population \citep[e.g.,][]{Botteon2020a, bruno23a}. Moreover, the spectral index of the cluster is comparable to that of radio halos in massive merging clusters \citep[e.g.,][]{Rajpurohit2021b, Bonafede2022}, indicating that the size, spectral, and power properties of the RXCJ0232 halo closely resemble those of the general giant radio halo population. The spatially resolved spectral index maps reveal a relatively uniform spectral index distribution across the halo, with values predominantly in the range $-1.3 \lesssim \alpha \lesssim -1.0$ and only local variations within the uncertainties. The lack of strong radial or azimuthal steepening indicates that particle acceleration is not confined to localised regions but instead operates over large, cluster-wide scales. 

Although uncommon, the presence of a giant radio halo in a relaxed cluster is not necessarily incompatible with the turbulent reacceleration scenario. In this framework, radio halos are primarily generated during merger events, with major mergers producing the classical population of bright halos. In contrast, minor mergers are expected to give rise to steep-spectrum halos \citep[e.g.,][]{Brunetti2007, Cassano2023, cuciti23}. Recent studies of Abell 1795 have shown that extended, Mpc-scale radio emission can coexist with a strong cool core, with the radio and X-ray morphologies closely tracing each other \citep{vanweeren2026}. This system also highlights that similar observational properties—such as flat spectral indices and large halo sizes—may arise in relaxed clusters. The timing of turbulence injection may also play a role, as a recent episode of turbulence can result in a relatively flat spectral index even when the total turbulent energy is very low. Likewise, a minor merger may generate a turbulent energy density comparable to that of a major merger if the energy is dissipated over a smaller volume. In this context, the relatively small characteristic radius, $r_{\mathrm{e}}$, compared to the average for the giant halo population, suggests that the radio halo in RXCJ0232 is approximately a factor of 3.3 less voluminous than a typical halo.

\subsection{Thermal and Nonthermal Correlation}

The $\rm I_{\rm R}–\rm I_{\rm X}$ spatial correlation analysis of RXCJ0232 reveals a consistently sublinear correlation at all observed frequencies (400, 650, and 1283 MHz), with no significant variation of the slope across the frequency range (Figure~\ref{fig: xray}). Such sublinear behaviour implies that the diffuse radio emission is more broadly distributed than the thermal X-ray emission, indicating slow spatial variations in the synchrotron emissivity. Although point-to-point studies of radio halos remain limited to a small number of well-resolved systems, previous work has reported both frequency-dependent and frequency-independent correlation slopes \citep[e.g.,][]{Govoni2001, Shimwell2014, Hoang2021, Rajpurohit2021a, santra24b}. In this context, RXCJ0232 belongs to the latter class, suggesting that the relativistic electron population responsible for emission across the observed frequency range is spatially well mixed, in line with the nondetection of radial spectral steepening on resolved scales.

In the context of particle acceleration models, sublinear $\rm I_{\rm R}$–$\rm I_{\rm X}$ correlations are naturally produced in turbulent reacceleration scenarios, where relativistic electrons are reenergised by volume-filling turbulence and the synchrotron emissivity depends on both the magnetic field distribution and the local turbulence properties. In contrast, hadronic models generally predict superlinear correlations, as the production of secondary electrons is closely tied to the thermal gas density. The observed sublinear slopes in RXCJ0232 therefore disfavour a purely hadronic origin as the dominant mechanism responsible for the radio halo emission. The lack of any measurable variation in the correlation slope across the 400–1283 MHz frequency range further supports a turbulent reacceleration scenario. Despite hosting a cool core and displaying an overall regular X-ray morphology, RXCJ0232 shows evidence of ICM perturbations, likely associated with a past minor merger or an asymmetric accretion event. In such a scenario, the merger-driven disturbance is insufficient to disrupt the cluster core but can inject turbulence into the outer regions of the ICM, providing the necessary conditions for particle reacceleration and the formation of a giant radio halo. This system represents a compelling example of a system in which turbulent reacceleration operates in a moderately disturbed environment, highlighting that major, core-disrupting mergers may not be a prerequisite for the generation of large-scale diffuse radio emission.

\section{Summary and Conclusions} \label{sec: summary}

We present a multiwavelength view of the cluster RXCJ0232. This cluster is one of the rare transitioning objects that shows signs of both GRHs and MHs. The peak of the radio emission coincides with the BCG (BCG-A), similar to MHs. However, the radio emission extends beyond 1 Mpc, exceeds the MHs, and is typical of GRHs. Our sensitive uGMRT observations ($\sim 300-750$ MHz), combined with available MeerKAT and Chandra X-ray observations, helped us obtain a detailed understanding of the spectral nature and thermal-nonthermal correlations in the ICM, providing physical insights into the origin of this enigmatic, transitioning radio halo. Our main findings are summarised below.

\begin{itemize}
    \item We present the first point-source-subtracted radio maps of the cluster RXCJ0232 in uGMRT bands 3 and 4, and the MeerKAT L band. The central extended radio halo emission is well recovered at all frequencies, with the largest linear extent of approximately 1.2 Mpc. We have also detected the candidate east relic with a linear size of approximately 300 kpc.

    \item We fitted the integrated flux densities of the radio halo and the candidate east relic with a single power law, yielding spectral indices of $-1.17 \pm 0.17$ and $-0.85 \pm 0.17$, respectively. We have also fitted the radio surface brightness profile for the halo with a simple exponential law. The e-folding radius of the radio profile shows no significant frequency variation. This implies that there is no radial spectral steepening over different frequencies.

    \item We created two resolved spectral index maps: one using uGMRT at 400 and 650 MHz, and the other using uGMRT at 400 MHz and MeerKAT at 1283 MHz. Both maps exhibit minimal variation in spectral indices across pixels. Most of the pixels have spectral indices around $-1.0$ to $-1.3$, which is not very steep for a radio halo with a cool core. This implies that in situ reacceleration of charged particles is occurring on small scales throughout the cluster region.

    \item  The point-to-point radio and X-ray surface brightness correlation analysis shows a strong positive correlation between the nonthermal and thermal parts of the ICM. The correlation slope is around $0.80$ and remains relatively stable across frequencies.

    \item The X-ray morphology and thermodynamic properties indicate that RXCJ0232 is an intermediate dynamical system, hosting a cool core while showing evidence for mild substructure. The coexistence of a cool core and an Mpc-scale radio halo suggests that turbulence generated by minor mergers or asymmetric accretion is sufficient to power cluster-wide particle reacceleration without disrupting the core.

\end{itemize}

All the above results, including the presence of two BCGs \citep{Pierini2008} and a faint substructure $\sim 1^{\prime}$ away from the cluster centre \citep{Parekh2021}, indicate that the cluster likely underwent a past merger event. The merger did not disturb the core of the cluster; however, it may have induced turbulence in the ICM, re-accelerating particles and powering the growth of the radio emission from a minihalo to a giant radio halo. The candidate east relic could have formed during this merging process, as its linear shape is perpendicular to the line joining the two BCGs. However, we do not have any strong evidence of this. Also, the distance of the candidate east relic from the cluster centre is significant ($\sim 1.05$ Mpc). Therefore, the emission could also be a remnant radio galaxy. Further work in the future, including additional deep X-ray and radio observations below 300 MHz, will provide a complete picture of the origin of this rare transitioning halo in the cluster RXCJ0232.

\section{Acknowledgments}
We thank the anonymous referee for the constructive comments that have improved the clarity of the paper. P.B. acknowledges Sameer Tanaji Salunkhe for useful discussions and his help in creating the multiwavelength composite image. P.B., R.S. and R.K. acknowledge the support of the Department of Atomic Energy, Government of India, under project No. 12-R\&D-TFR-5.02-0700. R.K. also acknowledges the support from the SERB Women Excellence Award WEA/2021/00000. We thank the staff of the GMRT that made these observations possible. The GMRT is run by the National Centre for Radio Astrophysics (NCRA) of the Tata Institute of Fundamental Research (TIFR). The scientific results reported in this article are based in part on data obtained from the Chandra Data Archive (Obs ID 4993). This research employs a list of Chandra datasets, obtained by the Chandra X-ray Observatory, contained in DOI:560 \footnote{\url{https://doi.org/10.25574/cdc.560}}. The MeerKAT telescope is operated by the South African Radio Astronomy Observatory, which is a facility of the National Research Foundation, an agency of the Department of Science and Innovation. This research has made use of NASA's Astrophysics Data System and of the NASA/IPAC Extragalactic Database (NED), which is operated by the Jet Propulsion Laboratory, California Institute of Technology, under contract with the National Aeronautics and Space Administration.

%

\vspace{5mm}
\facilities{uGMRT (upgraded Giant Metrewave Radio Telescope), MeerKAT (Meer Karoo Array Telescope), CXO (Chandra X-ray Observatory).}


\software{CASA \citep{casa}, astropy \citep{Astropy2013, Astropy2018, astropy2022}, numpy \citep{numpy}, aplpy \citep{aplpy1, aplpy2}, matplotlib \citep{matplotlib}, ds9 \citep{ds91, ds92, ds93, ds94, ds95}.}



\appendix \label{appn}

\section{Flux Offset in the Baselines of the GMRT band--4 Central Square} \label{issue}

\begin{figure}
    \centering
    \includegraphics[width = 0.70\textwidth]{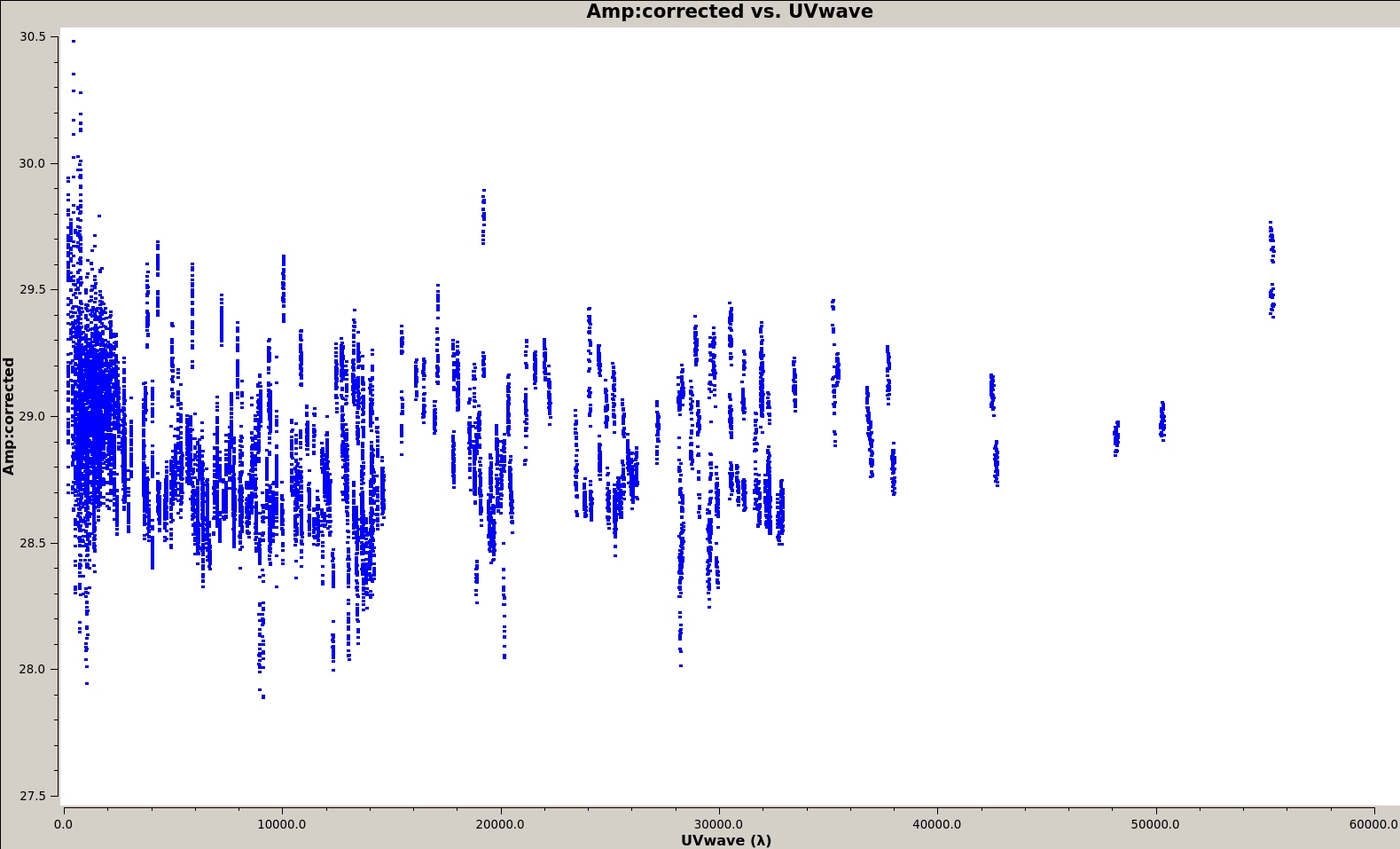}
    \caption{Calibrated amplitude versus uv-distance (UVwave) for the flux calibrator 3C48 in band 4 of the uGMRT. The plot shows a systematic flux density offset between the shorter and longer baselines of the GWB data during the 2019 period. This issue necessitated heavy flagging of the short-baseline data, leading to the reduced flux recovery for the diffuse emission at 650 MHz.}
    \label{fig: offset}
\end{figure}

The GMRT array comprises 30 fully steerable $45$ m diameter parabolic dishes in a hybrid configuration.  It consists of 14 antennas (labelled C00--C14) in a compact, randomised ``central square" ($\sim \rm 1 \ km \times 1 \ km$) and 16 antennas along three arms in Y-shaped (labelled east, west, south), extending up to 25 km. During 2019, GMRT experienced flux density offsets in the central square antenna baselines with the GWB\footnote{\url{https://www.gmrt.ncra.tifr.res.in/gmrt_users/help/csq_baselines.html}}. This issue was most severe in band 5, moderate in band 4, and minimal in band 3, with baselines between the central square and arm antennas, or between arm antennas, being less affected. Our band 3 data show no offset, but band 4 data have an offset (Figure \ref{fig: offset}). The band 4 visibilities also exhibit ripple-like artefacts (slight oscillation in the amplitudes) at longer baselines, possibly because of this offset issue. Using the known flux densities of 3C48 at various frequencies, we estimated its 650 MHz flux density to be $\sim28.56$ Jy. However, short baselines yielded a higher value ($\sim29$ Jy), confirming the offset. Consequently, most central square baseline data were flagged, preventing detection of the full halo extent in band 4 (Figure \ref{fig: diffuse}) and leading to a loss of flux density (Figure \ref{fig: intx}).

\section{Resolved Spectral Index Map between the GMRT Bands 3 and 4} \label{rsindxGMRT}

\begin{figure*}
    \centering
    \includegraphics[width = 1\textwidth]{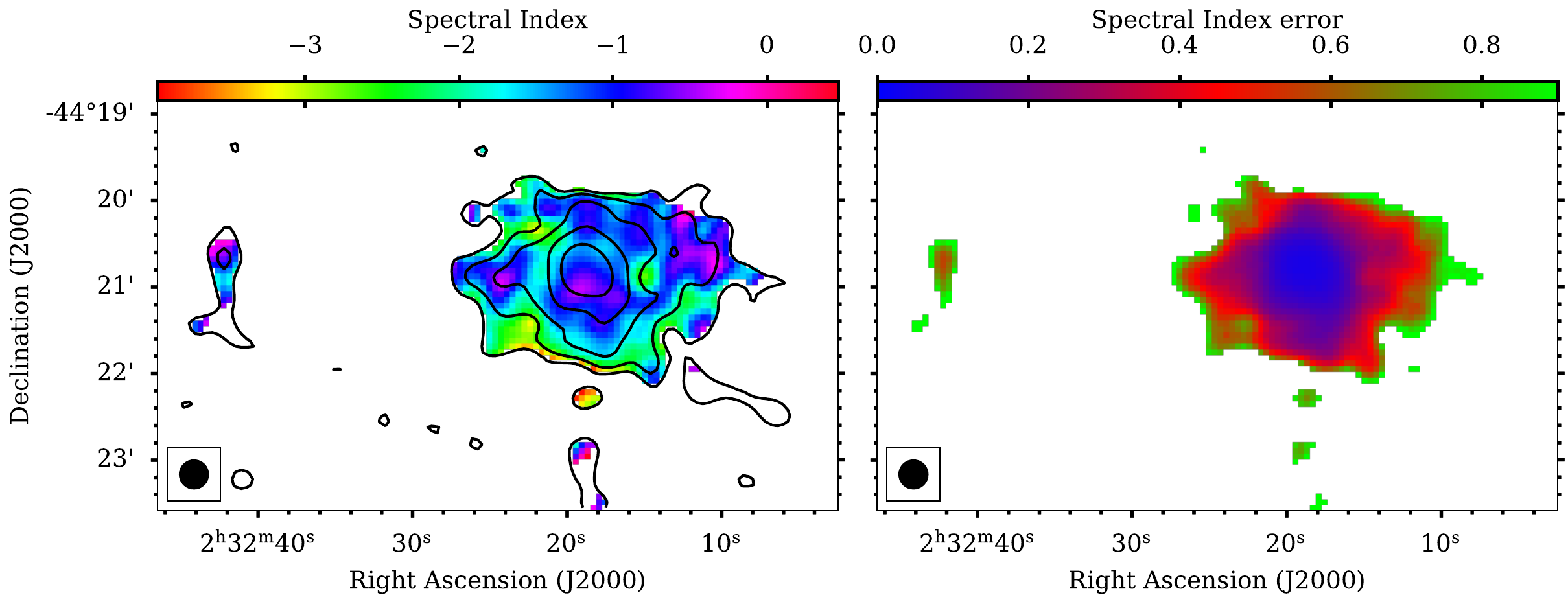}
    \caption{Spatially resolved spectral index map (\textit{left}) of the central region of cluster RXCJ0232 at $20^{\prime\prime}$ resolution, calculated between uGMRT 400 MHz and 650 MHz, with the corresponding error map shown on the \textit{right}. Black contours represent the uGMRT 650 MHz emission with levels at $[1, 2, 4, 8, \dots] \times 3\sigma_{\text{rms}}$, where $\sigma_{\text{rms}} = 56$ $\mu$Jy beam$^{-1}$.}
    \label{fig: rindxgmrt}
\end{figure*}

Because of the limitation (see appendix \ref{issue}) in the band 4 data of the uGMRT, we could not recover the full flux density and extent of the radio halo. The resolved spectral index map between the uGMRT bands 3 and 4 is shown in Figure \ref{fig: rindxgmrt}.

\bibliography{sample631}{}
\bibliographystyle{aasjournal}



\end{document}